\documentstyle[prl,aps,twocolumn]{revtex}

\def\5{\bar }  \def\6{\partial } \def\7{\tilde }
\def\8{\hat }
\def\hmu{\hat{\mu}}
\def\hnu{\hat{\nu}}
\def\F{{\cal F}} \def\G{{\cal G}}

\begin{document}

\title{
T-Duality and Effective D-Brane Actions
}

\author{
Joan Sim\'on}

\address{
Departament ECM,
Facultat de F\'{\i}sica,
Universitat de Barcelona and 
Institut de F\'{\i}sica d'Altes Energies,
Diagonal 647,
E-08028 Barcelona, Spain.
E-mail: jsimon@ecm.ub.es.\\[1.5ex]
\begin{minipage}{14cm}\rm\quad
T-duality realized on D-brane effective actions is studied from a pure
worldvolume point of view. It is proved that invariance in the form of
the Dirac-Born-Infeld and Wess-Zumino terms fixes 
the T-duality transformations of the NS-NS and R-R background fields,
respectively. The analysis is extended to uncover the mapping of global
symmetries of the corresponding pair of D-branes involved in the
transformation.\\[1ex]
PACS numbers: 11.10.Kk, 11.25.-w\\
Keywords: T-duality, D-branes, Dirac-Born-Infeld actions, global symmetries
\end{minipage}
}

\maketitle

%%%%%%%%%%%%%% body of paper %%%%%%%%%%%%%%%%%%%%%%%%%

T-duality is known to be a perturbative symmetry of string theory ~\cite{Kik}.
It states that a string compactified on a torus with spacelike radius $R$
and string coupling constant $g$ is equivalent to a string compactified on a 
torus with spacelike radius $\alpha^{\prime}/R$ and string coupling constant
$g\,\sqrt{\alpha^{\prime}}/R$. It is believed to be an exact symmetry of the 
whole string theory, that is, including the non-perturbative sector. 

{}One of the evidences in favour of this conjecture is the realization of 
T-duality on Dirichlet $p$-brane effective actions ($Dp$-branes) 
~\cite{eric1}. The final conclusion in ~\cite{eric1} was that the double 
dimensional reduction of the $Dp$-brane action yields the direct dimensional 
reduction of the $D(p-1)$-brane action. This approach was based on the 
T-duality rules mapping type IIA/IIB backgrounds given in ~\cite{eric},
which were derived due to the existence of two inequivalent
embeddings of $N=2$ $D=9$ supergravity into the corresponding ten dimensional
ones. 

{}The derivation of the T-duality mapping rules for the NS-NS sector using
pure worldvolume techniques was achieved in ~\cite{ver}, where they gauged 
the compact isometry in a non-linear sigma model describing the propagation 
of a fundamental string in a curved background. It seems rather natural to 
consider, instead, $Dp$-brane effective actions, because not only they couple 
to the NS-NS sector but also minimally to the whole R-R sector.

{}$Dp$-branes admit a CFT description within open string theory ~\cite{ref2}.
The dimensionality of their worldvolume depends on the number of scalar 
fields $(9-p)$ satisfying Dirichlet boundary conditions (b. c.). Since
Dirichlet b. c. are transformed to Neumann b. c. under T-duality  ~\cite{pol},
$D(p+1)$-branes are T-dual to $Dp$-branes. Since the dynamics of all 
$Dp$-branes is described by,

\begin{eqnarray}
S_{p} & = & -T_p \int d^{p+1}\sigma \,e^{-\phi}\,\sqrt{-det(\G_{\mu\nu} 
+ \F_{\mu\nu})} \nonumber \\
& & + T_p \int e^{\F}\wedge C
\end{eqnarray}

\noindent where $\G_{\mu\nu} \equiv \6_\mu x^M \6_\nu x^N g_{MN}$ is the
worldvolume induced metric and 
$\F_{\mu\nu} \equiv 2\pi\alpha^{\prime}F_{\mu\nu} - 
\6_\mu x^M \6_\nu x^N B_{MN}$, the above discussion suggests that T-duality 
rules mapping type IIA/IIB supergravity backgrounds must be encoded in the 
Dirac-Born-Infeld term (DBI), for the NS-NS sector, and in the Wess-Zumino 
term (WZ) for the R-R sector.

{}It will be proven that requiring invariance in the form of the DBI and WZ
terms fixes the T-duality transformations of the NS-NS and R-R background
fields, respectively. In particular, the complete set of transformations for 
the R-R potentials will be determined for an arbitrary target space background,
generalizing previous results in the literature. The methods developed
in this proof might be useful to treat the kappa symmetric case, in 
order to derive the mapping rules of the fermionic supergravity sector,
and to realize timelike and spacetime signature changing T-duality 
transformations ~\cite{hull} on exotic D-brane worldvolume effective actions.

{}The mentioned analysis refers to effective actions. It is extended to uncover
the global symmetries of these effective actions. The mapping relating the 
original infinitessimal transformations with the T-dual ones is found under 
certain sufficient conditions. We comment on the physical interpretation
of this result.
 
%%%%%%%%%%%%%%%%%%%%%%%%%%%%%%%%%%%%%%%%%%%%%%%%%%%%%

{\em T-duality from Born-Infeld invariance .}
%\label{second}
Consider a $Dp$-brane propagating in an arbitrary type II supergravity
background. Its bosonic degrees of freedom are $\{\phi^i\} = \{x^M,V_{\mu}\}$,
where $x^M$ are the target space coordinates, $M =0,1 \ldots 9$ and 
$V_{\mu}$ are the components of the 1-form worldvolume gauge field, $\mu =0,1
\ldots p$. 

{}Given the field content, one must specify the set of global and gauge 
symmetries that leaves the corresponding action invariant. Let us write down 
the infinitessimal transformations of the fields 
$\{\phi^i\}$ associated with both types of symmetries :

\begin{eqnarray}
s\,x^M & = & \xi^{\nu}\6_{\nu}x^M + \Delta x^M 
\label{trafo1} \\
s\,V_\mu & = & \xi^{\nu}\6_{\nu}V_\mu + V_\nu\6_\mu \xi^\nu + \6_\mu c +
\Delta V_\mu
\label{trafo2}
\end{eqnarray} 

\noindent where $s$ is a generalized worldvolume BRST operator, $\xi^\nu$ are 
diffeomorphism ghosts, $c$ an abelian $U(1)$ ghost, and $\Delta\phi^i$ stands 
for the infinitessimal set of independent global transformations.

{}All $Dp$-branes possess the above structure, differing just in the
supergravity background (IIA or IIB) and in the worldvolume dimension. The
$D(p+1)$-brane effective action can be mapped to the $Dp$-brane one, if
one dynamical component of the initial gauge field is identified with
an scalar in the T-dual $Dp$-brane, as a necessary condition to match
the different physical degrees of freedom described by both $D$-branes.

{}The latter discussion suggests demanding the conditions
of a double dimensional reduction, as the worldvolume dimension
must be effectively reduced by one. This mechanism is equivalent
to a partial gauge fixing of the worldvolume diffeomorphisms $(z = \rho)$
combined with a functional truncation $\6_\rho x^m =\6_\rho V_\mu = 0$,
where the original $x^M$ scalar fields and the worldvolume coordinates 
$\sigma^\mu$ were splitted into $\{x^m,z\}$ and $\{\sigma^{\8\mu},\rho\}$,
respectively. 

{}In order to apply the above procedure, it is necessary to work with a 
lagrangian density independent of the $z$ coordinate in the subspace of
field configurations defined by the truncation. This is ensured by assuming the
existence of an isometry along the $z$ direction in the background where the
$D(p+1)$-brane is propagating, that is there exists $\Delta z =\epsilon\, ,\,
\Delta x^m = 0 \, , \,\Delta V_\mu = 0$ among the set of global symmetries
$\{\Delta\phi^i\}$.

{}If the functional truncation is extended to the ghosts 
$\xi^{\8\mu}$ and the space of rigid transformations is 
restricted to the subspace defined by $\6_z \Delta x^M = \6_z \Delta 
V_\mu = 0$, the consistency conditions requiring the infinitessimal 
transformations not to move our field configurations from the subspace 
defined by the truncation and partial gauge fixing $(\6_z s\phi^i 
\mid_{g.f. + trunc}=0)$, determines

\begin{equation}
c(\sigma^{\8\mu},\rho) = \7c (\sigma^{\8\mu}) + a + {\epsilon^{\prime}
\over 2\pi\alpha^{\prime}}\rho
\end{equation}

\noindent where $a,\epsilon^{\prime}$ are constants, and $\epsilon^{\prime}$
has mass dimension minus one. The corresponding transformations that 
leave invariant the partially gauge fixed truncated action become

\begin{eqnarray}
\7s x^m & = & \xi^{\8\nu}\6_{\8\nu}x^m + \7\Delta x^m
\label{trafo11} \\
\7s V_{\8\mu} & = & \xi^{\8\nu}\6_{\8\nu}V_{\8\mu} + V_{\8\nu}\6_{\8\mu}
\xi^{\8\nu} + \6_{\8\mu}\7c + \7\Delta V_{\8\mu}
\label{trafo21} \\
\7s V_\rho & = & \xi^{\8\nu}\6_{\8\nu}V_\rho + \7\Delta V_\rho 
\label{trafo22}
\end{eqnarray}

\noindent where $\7\Delta V_{\8\mu} = \Delta V_{\8\mu} - 
V_\rho\6_{\8\mu} \Delta z$, $\7\Delta V_\rho = \Delta V_\rho + 
{\epsilon^{\prime} \over 2\pi\alpha^{\prime}}$ and $\7\Delta x^m$ satisfies
$\6_z \7\Delta x^m = 0$.

{}From (\ref{trafo22}), not only $V_\rho$ becomes a scalar for the
partially gauge fixed action, which will be identified with a new
target space coordinate (defined in the T-dual spacetime), 
$(2\pi\alpha')\,V_\rho \equiv \7z$, but also the isometry along the 
$\7z$ direction is generated as a residue of the original $U(1)$ gauge 
symmetry $(\epsilon^{\prime})$ 
\footnote{
Here, the sufficient conditions have been given to derive this
isometry, but a more general analysis would be interesting to be
performed, due to the big symmetry structure of these effective
actions.}. 

{}Once the matching between bosonic degrees of freedom has been done, we
move to prove that invariance of the Dirac-Born-Infeld term fixes
the T-duality transformations of the NS-NS sector. Working on the partially
gauge fixed truncated subspace of configurations, the following identities
can be derived :

\begin{eqnarray}
\G_{\hmu \rho} & = & \6_{\hmu} x^m \5g_{mz} \\
\G_{\rho\rho} & = & \5g_{zz} \\
\G_{\hmu\hnu} & = & \G'_{\hmu\hnu} + \6_{\hmu}x^m \6_{\hnu}x^n (\5g_{mn} -
g'_{mn}) - \6_{\hmu}x^m \6_{\hnu}\7z g'_{\7zm} \nonumber \\
& & - \6_{\hmu}\7z \6_{\hnu}x^{\7M}g'_{\7z\7M} \\
\F_{\hmu \rho} & = & \6_{\hmu}\7z - \6_{\hmu}x^m\5B_{mz} \\
\F_{\hmu \hnu} & = & \F'_{\hmu \hnu} -\6_{\hmu}x^m\6_{\hnu}x^n(\5B_{mn}-
B'_{mn}) + \6_{\hmu}\7z\6_{\hnu}x^n B'_{\7zn} \nonumber \\
& & + \6_{\hmu}x^m\6_{\hnu}\7z B'_{m\7z}
\end{eqnarray}

\noindent where barred fields refer to our original background and
primed fields to dual backgrounds, and $\{\7M\}=\{m,\7z\}$ are the indices
of the dual target space.

{}The elements of the matrix whose determinant appears in the kinetic term 
of the Born-Infeld action can be computed from the above identities. Its
determinant is equal to

\begin{eqnarray}
& & det\,(\G_{\mu\nu}+\F_{\mu \nu}) = 
\5g_{zz}\,det\,[\G'_{\hmu\hnu}+
\F'_{\hmu \hnu} \nonumber \\
& & +\6_{\hmu}x^m\6_{\hnu}x^n\,[(\5g_{mn}-g'_{mn}) - (\5B_{mn}-B'_{mn})
\nonumber \\
& & -(\5g_{mz}-\5B_{mz})(\5g_{nz}+\5B_{nz})(\5g_{zz})^{-1}] \nonumber \\
& & -\6_{\hmu}x^m\6_{\hnu}\7z\,[(g'_{m\7z}-B'_{m\7z}) - (\5g_{mz}-\5B_{mz})
(\5g_{zz})^{-1}] \nonumber \\
& & -\6_{\hmu}\7z\6_{\hnu}x^n\,[(g'_{\7zn}+B'_{n\7z}) + (\5g_{nz}+\5B_{nz})
(\5g_{zz})^{-1}] \nonumber \\
& & -\6_{\hmu}\7z\6_{\hnu}\7z\,(g'_{\7z\7z}-\frac{1}{\5g_{zz}})]
\end{eqnarray}

{}We will require the DBI action to equal 
$-T_{p-1}\int d^p \sigma\, e^{-\phi'}\,\sqrt{-det\,(\G'_{\hmu\hnu}+
\F'_{\hmu \hnu})}$. Since the starting DBI lagrangian 
density was assumed to be independent of $\rho$,
the integration over this direction can be performed, so that
$T_{p-1} = T_p \, L_z $, which is the usual equation relating two D-brane 
tensions connected by a longitudinal T-duality transformation. From the 
lagrangian density itself, we get a set of six sufficient constraints, which 
are obtained by functional independence and taking symmetric and antisymmetric 
parts when necessary. Its solution, given below, coincides with the usual 
T-duality rules mapping the bosonic NS-NS sector of type IIA supergravity 
to IIB or viceversa:

\begin{eqnarray}
\5g_{zz} & = & 1/g'_{\7z\7z} \nonumber \\
\5\phi & = & \phi' - \frac{1}{2}log\mid g'_{\7z\7z}\mid \nonumber \\
\5B_{nz} & = & -g'_{n\7z}/g'_{\7z\7z} \nonumber \\
\5g_{nz} & = & -B'_{n\7z}/g'_{\7z\7z} \nonumber \\
\5g_{mn} & = & g'_{mn} - (g'_{m\7z}g'_{n\7z}-B'_{m\7z}B'_{n\7z})/g'_{\7z\7z} 
\nonumber \\
\5B_{mn} & = & B'_{mn} - (B'_{m\7z}g'_{n\7z}-B'_{n\7z}g'_{m\7z})/g'_{\7z\7z}
\label{Tduality}
\end{eqnarray}

{}Let us concentrate on the R-R sector. The Wess-Zumino lagrangian density 
satisfies ${\cal L}_{WZ} = d\rho\wedge i_{\6{_\rho}} {\cal L}_{WZ}$,
which can be rewritten as

\begin{equation}
{\cal L}_{WZ} = d\rho\wedge e^{\F^-}\wedge\left(i_{\6_{\rho}} C + 
i_{\6_{\rho}} \F \wedge C^{-}\right) 
\label{WZ}
\end{equation}

\noindent where $\F^- \equiv i_{\6_{\rho}}(d\rho\wedge \F)$,
$i_{\6_{\rho}}\Omega_{(n)} = {1 \over (n-1)!} 
\Omega_{\rho\mu_2 \ldots \mu_n} d\sigma^{\mu_2}\wedge \ldots d\sigma^{\mu_n}$
and $i_{\6_{\rho}} (\Omega_{(m)}\wedge \Omega_{(n)}) =  i_{\6_{\rho}}
\Omega_{(m)}\wedge \Omega_{(n)} + (-1)^m \Omega_{(m)}\wedge 
i_{\6_{\rho}}\Omega_{(n)}$.

{}As we already know the T-duality properties of the two form $\F$, which
were fixed by the kinetic term,

\begin{eqnarray}
& T : \quad \F^- \longrightarrow \F' - i_{\6_{\7z}} B' \wedge i_{\6_{\7z}} g' 
/g'_{\7z\7z} & 
\label{FT1} \\
& T: \quad i_{\6_{\rho}} \F \longrightarrow - i_{\6_{\7z}} g' 
/g'_{\7z\7z} &
\label{FT2}
\end{eqnarray}

\noindent we can apply a T-duality transformation to (\ref{WZ}),  and demand
its transformed to equal $T_{p-1}\int_{\6 M} e^{\F'}\wedge C'$. In this way,
not only $T_{p-1} = T_p L_z$ is satisfied, but the following condition is 
obtained for the pullback of an arbitrary p-form :

\begin{eqnarray}
(-1)^pC'_{(p)} & = & i_{\6_{\rho}}\overline{C}_{(p+1)} - {i_{\6_{\7z}} 
B'\wedge i_{\6_{\7z}} g' \over g'_{\7z\7z}}\wedge i_{\6_{\rho}}
\overline{C}_{(p-1)} \nonumber \\
& & - {i_{\6_{\7z}} g' \over g'_{\7z\7z}}\wedge\overline{C}^{-}_{(p-1)}\, .
\label{RR}
\end{eqnarray}

\noindent where the factor $(-1)^p$ is due to our conventions 
combined with $\epsilon^{\7\mu_1 \ldots \7\mu_p} \equiv
\epsilon^{\mu_1 \ldots \mu_p \rho}$ and $\epsilon^{01 \ldots p}=1$. 

{}In view of the relation $dz=d\rho$, the $\pm$ components of the pullbacked 
worldvolume forms appearing in (\ref{RR}) can be lifted to $\pm$ components
of the space-time forms. Its corresponding solution is given by the following 
pair of expressions :

\begin{eqnarray}
i_{\6_{z}}\overline{C}_{(p+1)} & = & (-1)^p \left(C'_{(p)} - {i_{\6_{\7z}} g' 
\over g'_{\7z\7z}}\wedge i_{\6_{\7z}} C'_{(p)}\right) 
\label{TRR1} \\
\overline{C}^{-}_{(p-1)} & = & (-1)^{(p-1)}\left(i_{\6_{\7z}} C'_{(p)} -
i_{\6_{\7z}} B' \wedge \left( C'_{(p-2)}\right.\right. \nonumber \\
& & \left.\left.- {i_{\6_{\7z}} g' \over g'_{\7z\7z}}\wedge i_{\6_{\7z}} 
C'_{(p-2)} \right)\right)
\label{TRR2}
\end{eqnarray}
 
\noindent which are the T-duality rules for the R-R sector. In components,
equations (\ref{TRR1}) and (\ref{TRR2}) read as follows :

\begin{eqnarray}
\overline{C}^{(p+1)}_{m_1 \ldots m_p z} & = & C'^{(p)}_{m_1 \ldots m_p} -
pC'^{(p)}_{[m_1 \ldots m_{p-1}\7z}{g'_{m_p] \7z} \over g'_{\7z\7z}} 
\label{comp} \\
\overline{C}^{(p)}_{m_1 \ldots m_p} & = & C'^{(p+1)}_{m_1 \ldots m_p \7z}
-pC'^{(p-1)}_{[m_1 \ldots m_{p-1}}B'_{m_p ]\7z} \nonumber \\
& & -p(p-1)C'^{(p-1)}_{[m_1 \ldots m_{p-2}\7z}B'_{m_{p-1}\7z}{g'_{m_p ]\7z}
\over g'_{\7z\7z}}
\label{comp1}
\end{eqnarray}

{}In the following, the relation between the above T-duality transformations 
and the pre-existing ones ~\cite{yolanda} will be clarified. In any field 
theory, one has always some freedom to redefine the set of fields and
couplings one is interested in. This is precisely what happens in D-brane
effective actions. The set of RR gauge potentials describing the coupling
of the brane to the background is not unique. In particular, there are two
bases of RR fields used in the literature : the one appearing in kappa
symmetric D-brane actions, which is the one used along the whole paper,
and a second one with very specific transformation properties under
S-duality transformations (denoted by a superindex S in the following).
For instance, the 4-form $C_{(4)}$ is not S self-dual, but transforms
to $C_{(4)} - C_{(2)}\wedge B_{(2)}$. It is $C^S_{(4)} = C_{(4)} -
{1\over 2}C_{(2)}\wedge B_{(2)}$, the one being S self-dual. The same 
happens for the 6-form, $C^S_{(6)} = C_{(6)} -{1\over 4}
C_{(2)}\wedge B_{(2)}\wedge B_{(2)}$, where $C^S_{(6)}$ forms an S-doublet
with the dual to the NS-NS 2-form $B_{(2)}$ ~\cite{yolanda}. It is
straightforward to verify the equivalence between equations 
(\ref{comp}), (\ref{comp1}) and the pre-existing T-duality transformations 
using the above redefinitions. Furthermore, one finds

\begin{eqnarray}
\overline{C}^S_{m_1 \ldots m_6} & = & C'_{m_1 \ldots m_6 \7z} -6
C'_{[m_1 \ldots m_5}{g'_{m_6 ]\7z} \over g'_{\7z\7z}} \nonumber \\
& & -45\left( C'_{[m_1} - C'_{\7z}{g'_{[m_1 \7z} \over g'_{\7z\7z}}\right)
B'_{m_2m_3}B'_{m_4 m_5}B'_{m_6]\7z} \nonumber \\
& & -45C'_{[m_1m_2\7z}B'_{m_3m_4}\left(B'_{m_5m_6} -4B'_{m_5\7z}
{g'_{m_6 ]\7z} \over g'_{\7z\7z}}\right) \nonumber \\
& & -30C'_{[m_1\ldots m_4 \7z}B'_{m_5\7z}{g'_{m_6 ]\7z} \over g'_{\7z\7z}}
\end{eqnarray}

\noindent which was not computed in the calculations done in ~\cite{yolanda}.

%%%%%%%%%%%%%%%%%%%%%%%%%%%%%%%%%%%%%%%%%%%%%%%%%%%%%%%

{\em T-duality and global symmetries .}
%\label{three}
The equations determining the whole set of non-trivial global symmetries
of D-$p$brane effective actions is not known for arbitrary p, although they
have been derived, using cohomological methods, for the bosonic D-string
$(p=1)$ in ~\cite{paper1}. It is nevertheless known that among this set
one can always find the isometries of the background for appropiate 
transformations of the gauge field ~\cite{paul2} :

\begin{equation}
\Delta x^M = k^M , \quad \Delta V_\mu = \lambda_M \6_\mu x^M
\label{gsym}
\end{equation}

\noindent where $k^M$ and $\lambda_M$ satisfy

\begin{eqnarray}
& {\cal L}_k g = {\cal L}_k \phi = 0 & \nonumber \\
& {\cal L}_k B = d\lambda & \nonumber \\
& {\cal L}_k C_{(p+1)} = d\omega_{(p)} - \omega_{(p-2)}\wedge dB &
\label{cond}
\end{eqnarray}

\noindent for some set $\{\omega_{(p)}\}$.

{} The purpose of the present section is to show how the symmetries
(\ref{trafo11}), (\ref{trafo21}) and (\ref{trafo22}) of the partially gauge
fixed truncated action described by the generalized worldvolume BRST
operator, when the global ones are restricted to solutions
of (\ref{cond}) are mapped under T-duality to symmetries of the T-dual
brane. It should be stressed that this is not the same as saying that all
symmetries of the longitudinal T-dual D-brane action are already contained
in the set of symmetries of the original D-brane.

{}It is trivial to see that
the gauge symmetries described by the BRST operator $(\xi^{\8\mu},\7c)$
are mapped to the corresponding gauge symmetries of the dual field theory
as it corresponds to any dimensional reduction. Concerning the global 
symmetries, it would be interesting to find the corresponding $k'^M$, 
$\lambda'_M$ and $\{\omega'_{(p)}\}$ for the dual
brane in the dual background, satisfying analogous equations to those
appearing in (\ref{cond}). Studying the different components of the equations
in (\ref{cond}), and using (\ref{Tduality}), one can map the initial
generalized Killing equations (\ref{cond}) to the necessary ones for the
dual brane if the following identifications are made :

\begin{eqnarray}
& k'^m = \7k^m , \quad k'^{\7z} = \7\lambda_{z} & \nonumber \\
& \lambda'_m = \7\lambda_m , \quad \lambda'_{\7z} = \7k^z
\label{sym1} \\
& \omega'_{(p)} = (-1)^{(p+1)}\left(i_{\6_z} \7\omega_{(p+1)} - {i_{\6_{\7z}}g' \over g'_{\7z\7z}} \wedge \7\omega^{-}_{(p-1)}\right. &
\nonumber \\
& \left. -i_{\6_{\7z}}B'\wedge {i_{\6_{\7z}}g' \over g'_{\7z\7z}} \wedge 
i_{\6_z} \7\omega_{(p-1)}\right) \quad ,&
\label{sym2}
\end{eqnarray}

\noindent where it should be stressed that $\{\7k^m,\7k^z,\7\lambda_m,
\7\lambda_z,\7\omega_{(p)}\}$ are the solutions to (\ref{cond}) subject to the
constraint $\6_z \Delta\phi^i = 0$ and (\ref{sym2}) is defined up to a total
derivative. It is essential for this procedure to work that 
${\cal L}_{\6_{\rho}}(e^\F \wedge \omega ) = 0$, which is satisfied by 
hypothesis.

{}To sum up, the infinitessimal transformations for the dual D-brane in the 
dual background are given by

\begin{eqnarray}
\7s x^m & = & \xi^{\8\nu}\6_{\8\nu}x^m + k'^m \label{s1} \\
\7s\7z & = & \xi^{\8\nu}\6_{\8\nu}\7z + k'^{\7z} \label{s2} \\
\7s V_{\8\mu} & = & \xi^{\8\nu}\6_{\8\nu}V_{\8\mu} + V_{\8\nu}\6_{\8\mu}
\xi^{\8\nu} + \6_{\8\mu}\7c + {\lambda'_M \over 2\pi\alpha^\prime}\6_{\8\mu}x^M
\label{s3}
\end{eqnarray}

\noindent A couple of remarks concerning the above set of transformations :
{\em {(i)}} the term involving $\epsilon'$ in (\ref{trafo22}) was absorbed into
$\7\lambda_z$, without loss of generality as the latter is mapped to
$k'^{\7z}$ by eq. (\ref{sym1}) and the dual background is independent
of the $\7z$ coordinate, and {\em {(ii)}} to derive (\ref{s3}), $\7c$ has 
been redefined without loss of generality.

{}To get a better understanding of the above result, let us consider a
$Dp$-brane propagating in a $Dp$-brane background. The algebra of global
symmetries contains $ISO(1,p)\times SO(9-p)$ generated by the corresponding
Killing vectors $\{k^m,k^z\}$ \footnote{ Here we use the notation used in the
rest of the paper. It should be understood that $\{k^m\} = \{k^\perp,
k^\parallel\}$, where $\parallel=0,1 \ldots p-1$, $\perp =p+1, \ldots ,9$
and $k^z=k^p$.}. If we now consider a longitudinal T-duality transformation,
the conditions that were demanded to the infinitessimal global transformations
, $\6_z \Delta\phi^i =0$, break the isometry group into $ISO(1,p-1)\times
SO(9-p)\times R$, generated by $\{\7k^m,\7k^z\}$, where the abelian factor
R is common of dimensional reductions. Since $B_{(2)}=0$, the one form
$\lambda_{(1)}$ in (\ref{cond}) is exact, $\lambda_{(1)}=d\beta$. Once more,
the consistency conditions breaking the isometry group, restrict the
function $\beta$ to be of the form $\beta=az +b+\7\beta(x^m)$. It is
precisely in this way how $\7\lambda_z=a$ generates the abelian isometry
for the dual background through (\ref{sym1}). The final isometry group is the
one of a delocalised $D(p-1)$-brane, which is the one obtained when using
(\ref{Tduality}), and defers from the usual isometry group of a $D(p-1)$-brane
classical supergravity solution, which is $ISO(1,p-1)\times SO(10-p)$.

{}In relation with the results presented in ~\cite{paper1,paper3}, where it
was shown that the D-string has an infinite number of global symmetries,
one can indeed conclude from the discussion presented above that the
$D0$-brane (particle) also admits an infinite number of them, as was already
pointed out in ~\cite{paper4}.

%%%%%%%%%%%%%%%%%%%%%%%%%%%%%%%%%%%%%%%%%%%%%%%%%%%%%%%

{\em Acknowledgements .}
JS would like to thank Joaquim Gomis and David Mateos for helpful discussions
and to Joaquim Gomis for suggesting him the problem and encouraging him.
This work has been partially supported by AEN98-0431 (CICYT) and 1998SGR 00026
(CIRIT).

%%%%%%%%%%%%%%%%%%%%%%%%%%%%%%%%%%%%%%%%%%%%%%%%%%%%%%%%%%%%%%%%%%

\end{document}